\documentclass[pra,twocolumn,aps,showpacs,preprintnumbers,amsmath,amssymb,superscriptaddress]{revtex4-1}

\usepackage{graphicx}
\usepackage{dcolumn}
\usepackage{bm}
\usepackage{amsmath}
\usepackage{physics}
\usepackage{xcolor}
\usepackage[colorlinks]{hyperref}

\usepackage{color}
\usepackage{soul}

\begin{document}


\title{Relativity of quantum states in entanglement swapping}

\author{Chris Nagele}
\affiliation{State Key Laboratory of Precision Spectroscopy, School of Physical and Material Sciences,East China Normal University, Shanghai 200062, China}
\affiliation{New York University Shanghai, 1555 Century Ave, Pudong, Shanghai 200122, China} 
\affiliation{Department of Physics, New York University, New York, NY 10003, USA}

\author{Ebubechukwu O. Ilo-Okeke}
\affiliation{New York University Shanghai, 1555 Century Ave, Pudong, Shanghai 200122, China} 
\affiliation{Department of Physics, School of Physical Sciences, Federal University of Technology,\\ P. M. B. 1526, Owerri, Imo State 460001, Nigeria}

\author{Peter P. Rohde}
\affiliation{Centre for Quantum Software \& Information (QSI), Faculty of Engineering \& Information Technology, University of Technology Sydney, NSW 2007, Australia}

\author{Jonathan P. Dowling}
\affiliation{Hearne Institute for Theoretical Physics and Department of Physics and Astronomy, Louisiana State University, Baton Rouge, Louisiana 70803, USA}
\affiliation{National Institute of Information and Communications Technology,
4-2-1, Nukui-Kitamachi, Koganei, Tokyo 184-8795, Japan}
\affiliation{NYU-ECNU Institute of Physics at NYU Shanghai, 3663 Zhongshan Road North, Shanghai 200062, China}
\affiliation{CAS-Alibaba Quantum Computing Laboratory, University of Science and Technology of China, Shanghai 201315, China.}

\author{Tim Byrnes}
\email{tim.byrnes@nyu.edu}
\affiliation{New York University Shanghai, 1555 Century Ave, Pudong, Shanghai 200122, China} 
\affiliation{State Key Laboratory of Precision Spectroscopy, School of Physical and Material Sciences,East China Normal University, Shanghai 200062, China}
\affiliation{NYU-ECNU Institute of Physics at NYU Shanghai, 3663 Zhongshan Road North, Shanghai 200062, China}
\affiliation{National Institute of Informatics, 2-1-2 Hitotsubashi, Chiyoda-ku, Tokyo 101-8430, Japan}
\affiliation{Department of Physics, New York University, New York, NY 10003, USA}

\date{\today}

\begin{abstract}
The entanglement swapping protocol is analyzed in a relativistic setting,  where shortly after the entanglement swapping is performed, a Bell test is performed. For an observer in the laboratory frame, a Bell violation is observed between the qubits with the swapped entanglement.  In a moving frame, the order of the measurements is reversed, and a Bell violation is observed even though the particles are not entangled, directly or indirectly, or at any point in time. Although the measurement results are identical, the wavefunctions for the two frames are starkly different --- one is entangled and the other is not. Furthermore, for boosts in a perpendicular direction, in the presence of decoherence, we show that the maximum Bell violation can occur across non-simultaneous points in time.  This is a signature of entanglement that is spread across both space and time, showing both non-local and non-simultaneous aspects of entanglement.   
\end{abstract}


\maketitle

\section{Introduction}

The nature of entanglement has always been a point of intrigue since the early days of quantum mechanics \cite{einstein1935can}. In the last few decades, advances in experimental techniques have been able to test directly the spooky action at a distance by demonstrating its effects at increasingly larger distances. Adapting terrestrial free-space methods \cite{ursin07,ma12}, entanglement distribution and quantum teleportation to distances over $ 1000 $ km have now been performed \cite{yin2017satellite,ren2017ground}, using space-based technology. Another fundamental test is to measure the bounds to the speed of influence due to entanglement \cite{salert2008,yin2013}. In such experiments it is advantageous to have widely separated and near-simultaneous measurements to ensure that the two events are outside of the light cone of influence of each other.  This allows one to close the locality loophole~\cite{pearle1970,clauser1974,handsteiner2017}, where conspiring parties may mimic results that are attributed to entanglement. 

Combining special relativity with quantum mechanics leads to peculiar results see e.g. \cite{rideout2012,peres2004}), such as from the fact that simultaneity is relative according to the observer's frame.  Suarez and Scarani noticed that, for near-simultaneous measurements of an entangled pair, it is possible to reverse the order of measurements by moving to a suitable frame \cite{suarez1997}. This inspired several experiments \cite{scarani2000, stefanov2002, zbinden2001, gisin2002} and recently it was shown that this paradox can be resolved by taking into account the uncertainty in the time of measurement  \cite{richardson2014}. These relativistic contexts have rekindled debate on the foundations of quantum mechanics. It is a strange fact of modern physics that non-relativistic quantum mechanics --- which is not constructed with relativity in mind at all  --- still gives consistent results with special relativity, such as the impossibility of superluminal communication due to the no-cloning theorem. Investigations of relativistic effects beyond the no-signaling principle have been made \cite{oreshkov2012,brukner2014},  focusing on quantum causality and possible applications of event-order swapping to quantum circuits. 

\begin{figure}
\centering
  \includegraphics[width=\linewidth]{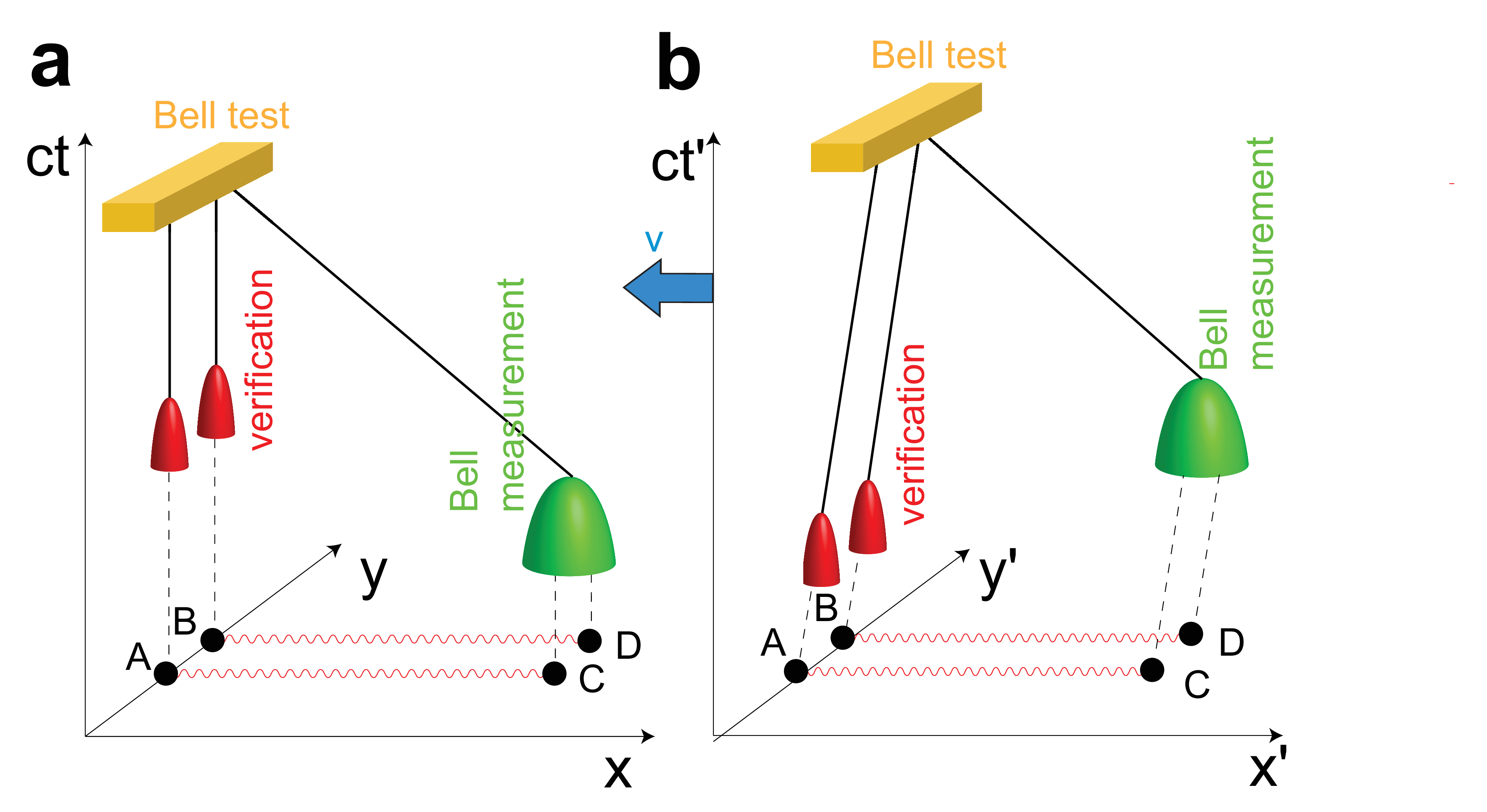}
\caption{Entanglement swapping illustrating the observer dependence of entanglement (Scenario 1).   Two reference frames   (a) the laboratory (Lara) with coordinates $(t,x,y,z) $; (b) an observer Rob with coordinates $(t',x',y',z') $, who is moving in the negative $ x $-direction are shown. Initially, four qubits are prepared, labeled by $ A, B, C, D $, where pairs $ AC $ and $BD $ are entangled (indicated by the wiggly lines). A Bell measurement and a verification measurement, consisting of two single-qubit measurements, are then made.  The order of the measurements depends upon the observer's frame.  In order to verify the Bell violation, the results of the Bell measurement must be classically transmitted to the verification measurements (or vice versa). Classical communication is denoted by thick solid lines, the dashed lines are to guide the eye.}
\label{fig1}
\end{figure}

In this paper, we examine the entanglement swapping protocol \cite{zukowski1993} in a relativistic setting, where, after entanglement has been swapped, a Bell test is used to verify the presence of the swapped entanglement.  We consider two scenarios in particular, which highlight two peculiar relativistic quantum effects. In Scenario 1, we contrast the interpretations of an experiment of two observers, which observe the Bell test and the entanglement swapping to take place in different orders (Fig. \ref{fig1}).  In the moving frame (Rob), we show that it is possible to observe a Bell violation, despite the fact that entanglement swapping has not occurred at all.  
In fact, Rob observes a Bell violation between two particles that has no entanglement whatsoever between them, directly or indirectly, or at any point in time.  This paradoxical effect occurs because of the quantum correlations, which --- even in the absence of entanglement--- determine the random measurement outcomes of the verification to give a Bell violation. 
In Scenario 2 depicted in Fig.~\ref{fig2}, we consider Rob to be moving in a perpendicular direction, and he is in control of the Bell test.  In this case, when there is some decoherence present, we find that there is an optimal time offset for him to make the two measurements to obtain the maximum Bell violation.  In this case the results of the Bell test point to correlations resulting from entanglement that are spread in both space {\it and} time.  

A similar experiment to Scenario 1 was considered by Peres \cite{peres2000delayed} in the context of delayed-choice entanglement swapping, and has been experimentally demonstrated since \cite{ma2012experimental,sciarrino2002delayed}.  Variations of the theme have been investigated, where entanglement can be verified between particles that have never co-existed \cite{megidish2013entanglement}. The difference of our scenario to these past works is that the order of the measurements depends upon the reference frame, due to relativity of simultaneity.  In particular, in Ref. \cite{megidish2013} the verification measurements and the Bell measurement are time-like separated. This procedure is complementary to, but differs from, our procedure where there are space-like separations between the Bell and verification measurements.  
Other works have investigated special relativistic effects on entanglement swapping \cite{beil2015, timpson2002} and on teleportation  \cite{friis2013, lin2015, nadeem2014, alsing2003}, but none have considered a setup where it is possible to switch the time ordering of measurements.  In relation to Scenario 2, several concepts relating to 
entanglement in time, and temporal Bell inequalities have been investigated \cite{legget1985,olson2012,budroni2013,zukowski2014}.  These differ from our scenario, since the nature of entanglement that they study is across time-like intervals and between the same system at different points in time.  In our case we consider entanglement in a more conventional sense, with a space-like orientation  and involving two particles. Specifically, the consecutive measurements on two different systems show stronger-than-classical correlations if the time difference between the two measurements matches the time it takes light to traverse the distance between the two systems as measured in a stationary frame.  

We note that throughout this paper we neglect the relativistic effects that occur locally on the quantum state, and assume that qubit measurements can be performed faithfully.  As has been discussed in several other works \cite{gingrich2002quantum,alsing2002lorentz,adami,byrnes2017lorentz}, depending upon the encoding of the logical qubit states, the local quantum state may be subject to Lorentz transformations.  For example, the polarization of a photon is not a Lorentz invariant quantity, and appears differently in different frames.  Since this is dependent upon the particular physical implementation, and can in principle be accounted for if the relative velocities are known, we disregard this effect and consider only the combined effect of quantum measurements with relativity of simultaneity.

\section{Scenario 1: Entanglement swapping}

\subsection{Laboratory frame (Lara)}
\label{sec:laraframe}

We now analyze the entanglement swapping protocol described in Fig. \ref{fig1}(a) in detail.  In this frame, the protocol appears as conventional entanglement swapping, where first a Bell measurement is made to swap the entanglement, then is followed by a verification measurement. Here, a Bell measurement is a measurement which projects qubits $ C$ and $D$ onto one of the four Bell states, while a verification measurement are made individually on $ A$ and $B $, projecting them onto a product state. 

 For concreteness, we consider the situation where all the measurements are stationary in Lara's frame. Qubits $ A $ and $ B $ are at the same $ x $-coordinate, and are separated from qubits $ C $ and $ D $ by a suitably large distance.  Initially, the qubits are prepared in the state 
\begin{align}
| \phi_0 \rangle = \ket{\Psi_{00}}_{AC} \ket{\Psi_{00}}_{BD}
\label{init_state}
\end{align}
where the four Bell states are defined as 
\begin{align}
\ket{\Psi_{ij}}  & = \frac{1}{\sqrt{2}}( \ket{0} \ket{i}+ (-1)^j \ket{1}\ket{1-i } ) \nonumber \\
& = I \otimes (\sigma^x)^i (\sigma^z)^j  \ket{\Psi_{00}},
\label{belldef}
\end{align}
for $ i, j \in \{0,1 \} $. These Bell states are equivalent to the traditional Bell states with $\ket{\Psi_{00}}=\ket{\Phi_{+}}, \; \ket{\Psi_{01}}=\ket{\Phi_{-}}, \; \ket{\Psi_{10}}=\ket{\Psi_{+}},$ and $\ket{\Psi_{11}}=\ket{\Psi_{-}}$.  After the Bell measurement is made the state collapses to one of the four outcomes with probability $ p_{ij} = 1/4 $
\begin{align}
\ket{\psi_{ij} }=\ket{\Psi_{ij}}_{AB}\ket{\Psi_{ij}}_{CD}.
\label{possible_states_a}
\end{align}
Soon afterwards (such that the events are space-like separated), a verification measurement is performed on qubits $A$ and $B$ to perform a Bell test. After the result of the Bell measurement has been classically transmitted, the appropriate CHSH quantity  \cite{clauser1969,aspect1981} is computed. The four states labeled by $ i, j $ are subject to their respective Bell tests and the result is averaged \cite{herbst2015, sun2017}.  Note that performing the Bell test separately on each of the four outcomes is necessary, otherwise the outcome would be a completely mixed state. For instance, the CHSH quantity for $\ket{\Psi_{00}}$
\begin{align}
S_{00} = \sum_{n=0}^1 \sum_{m=0}^1 \bra{\psi_{00}} (-1)^{(1-n)m} \hat{\mathcal{A}}_n \hat{\mathcal{B}}_{m}  \ket{\psi_{00} },
\label{CHSH_00}
\end{align}
is evaluated where the operators for qubits $ A $ and $ B $ are 
\begin{align}
\hat{\mathcal{A}}_n & = \cos (n \pi/2) \sigma_A^z + \sin (n \pi/2)  \sigma_A^x \nonumber \\
\hat{\mathcal{B}}_m & = \frac{\sigma_B^x +(-1)^m \sigma_B^z}{\sqrt{2}} 
\label{verifying}
\end{align}
for $ n,m \in \{0,1 \} $ respectively. This a group of four measurements --- only one of which is performed for a particular run of the experiment. The results of the verification measurement are used to compute the CHSH quantity for a general state $\ket{\psi_{ij}}$, which can be written (see Appendix)
\begin{align}
S_{ij} = \sum_{n=0}^1 \sum_{m=0}^1 \bra{\psi_{ij}} (-1)^{(1-n)m+j} \hat{\mathcal{A}}_n \hat{\mathcal{B}}_{m+i+j}  \ket{\psi_{ij} } .
\label{CHSH}
\end{align}
Each state observes a positive maximal Bell violation and the average is also a maximal Bell violation. We note that none of the verification data that is taken is thrown away --- in this sense no post-selection is performed. The data is however conditionally processed according to the appropriate CHSH quantity (\ref{CHSH}) for the the Bell state $\ket{\Psi_{ij}}$ in question.


%
%

\subsection{Rob's frame with $x$-boosts}

Now consider the moving frame of Rob, who is moving in the $-x$ direction. Due to the Bell measurement and the verification measurements being space-like separated, Rob observes that the latter occurs before the former [Fig. \ref{fig1}(b)]. Starting from (\ref{init_state}), the verification measurement yields 
\begin{align}
\bra{\phi_0} \hat{\mathcal{A}}_n \hat{\mathcal{B}}_m \ket{\phi_0} = 0
\label{noviolation}
\end{align}
Substituting this into (\ref{CHSH}) of course gives $S= 0$, which does not violate the Bell inequality, as expected for a product state between $ A $ and $ B $.  

The state collapses in the basis of the verifying measurements (\ref{verifying}).  For measurement settings $n,m \in \{0,1\}$ and measurement outcomes $l_A,l_B \in \{0,1\}$, the state after the measurement is
\begin{align}
 | l_A \rangle_A^{(n)}  | l_B \rangle_B^{(m)}  | l_A \rangle_C^{(n)}  |  l_B \rangle_D^{(m)} ,
\label{intermediatestaterob}
\end{align}
which occur with probabilities 
\begin{align}
p_{l_A l_B}^{(nm)}  = | \langle \Psi_{00} | l_A \rangle_A^{(n)} | l_A \rangle_C^{(n)} |^2 | \langle   \Psi_{00} |  l_B \rangle_B^{(m)}  |  l_B \rangle_D^{(m)} |^2  = \frac{1}{4}
\label{randomdata}
\end{align}
Here the states are the eigenstates of  (\ref{verifying})
\begin{align}
\hat{\mathcal{A}}_n | l \rangle^{(n)} & = (-1)^l |  l \rangle^{(n)} \nonumber \\
\hat{\mathcal{B}}_m | l \rangle^{(m)} & = (-1)^l |  l \rangle^{(m)} .
\end{align}
The Bell measurement then projects the $CD$ qubits onto the Bell states (\ref{belldef}).  If the Bell measurement returns $ | \Psi_{ij} \rangle $, then the obtained state is
\begin{align}
|  l_A \rangle_A^{(n)} | l_B \rangle_B^{(m)} | \Psi_{ij} \rangle_{CD}
\end{align}
which occurs with probability 
\begin{align}
p_{ij l_A l_B}^{(nm)} & = p_{l_A l_B}^{(nm)} | \langle \Psi_{ij} | l_A \rangle^{(n)}_C | l_B \rangle^{(m)}_D |^2  \nonumber \\
& = \frac{1}{16} \left( 1 + \frac{(-1)^{l_A - l_B + i (1-n) + jn + n(1-m) }}{\sqrt{2}}  \right),
\label{robprobs}
\end{align}
for the outcome $ (i,j,l_A,l_B ) $.  Now as before, we evaluate the CHSH correlation (\ref{CHSH}) for each of the $ i, j $ outcomes.  
Then the average value conditioned on the outcome $ (i,j) $ is
\begin{align}
\langle \hat{\mathcal{A}}_n \hat{\mathcal{B}}_m \rangle \Big|_{ij} & = \frac{\sum_{l_A l_B} (-1)^{l_A+l_B} p_{ij l_A l_B }^{(nm)}}{\sum_{l_A l_B} p_{ij l_A l_B }^{(nm)}} \nonumber \\ 
& = \frac{ (-1)^{(1-n)m} (-1)^{i(1-n) + nj}  }{\sqrt{2}}.
\label{abcorrrob}
\end{align}
Now, putting the expectation value in the same form as (\ref{CHSH}), we obtain
\begin{align}
\langle \hat{\mathcal{A}}_n \hat{\mathcal{B}}_{m+i+j} \rangle \Big|_{ij} = \frac{ (-1)^{(1-n)m+j}  }{\sqrt{2}}.
\label{robresult}
\end{align}
Substituting this into (\ref{CHSH}), this yields a Bell violation $S_{ij}=2\sqrt{2}$ for all $ i,j $. Averaging over all $ i,j $ of course again gives a Bell violation, giving the same result as Lara's frame.

\subsection{Relativity of entanglement}

We have seen that, ultimately, the results of the Bell test agree in both frames.  At one level, this is not surprising since it is a relativistic principle that measurement outcomes  must agree in all frames.  
Furthermore, the only role of the boosted frame is to reverse the order of the measurements.  Since the two types of measurements commute (they are on separate qubits, $AB $ and $ CD $), the fact that the same outcomes are obtained is always guaranteed \cite{peres2000delayed,peres2000classical,brukner2005complementarity,florig1997statistical}. 

What is unusual is that the interpretation of the experiment is completely different in the two frames.  Between the two measurements, the quantum state is different, due to the differing order of the measurements.  In Lara's frame, the intermediate state is (\ref{possible_states_a}), which is an entangled state between qubits $ AB $. This appears as entanglement swapping from the initial configuration to between $AB $ and $CD$. Using a conditional Bell test which depends upon the result of the Bell measurement, a Bell violation is observed.  Thus the origin of the Bell violation can be straightforwardly attributed to the presence of entanglement between $ AB$.  

On the other hand, the intermediate state in Rob's frame is (\ref{intermediatestaterob}), which has no entanglement at all.  In fact, in Rob's frame there is never any entanglement between $ A B $ from the beginning to the end of the sequence. The Bell violation appears to be more akin to manipulation of random data such as to violate a Bell inequality, since it is achieved by communication of appropriate side information by the Bell measurement on $CD$ and the outcomes at $AB$ are completely random by  (\ref{randomdata}).  

To see this point, let us compare Rob's point of view to a parallel thought experiment where there are only qubits at $ A $ and $ B $. The same verification measurements are performed on each qubit, and a conditional Bell test is performed.  In this case, first prepare a completely mixed state at $ AB$:
\begin{align}
\rho_{AB} = \frac{I_A \otimes I_B}{4}  ,
\label{completelymixedstate}
\end{align}
where $ I_A, I_B $ are $ 2 \times 2 $ identity matrices on $ A, B $. The measurement basis $ n,m $ of the verification are randomly chosen in the same way as (\ref{verifying}). Since the initial state is a completely mixed state, one obtains the outcomes
\begin{align}
 | l_A \rangle_A^{(n)}  | l_B \rangle_B^{(m)} 
\end{align}
with probability $ p_{l_A l_B}^{(nm)} = 1/4 $.  Now suppose there is a demon, who has knowledge of the particular measurement choices $ n, m $ and the outcomes $ l_A, l_B $.  The demon can then draw random variables $ i,j $ with probability 
\begin{align}
p^{(nm)}_{ij|l_A l_B} = \frac{1}{4} \left( 1 + \frac{(-1)^{l_A - l_B + i (1-n) + jn + n(1-m) }}{\sqrt{2}}  \right) .
\label{demonprob}
\end{align}
The demon then communicates the result to the Bell tester. We note the demon does not have to perform a Bell measurement to produce this probability distribution, he may simply follow the formula in (\ref{demonprob}).  The Bell tester then evaluates the corresponding CHSH inequality (\ref{CHSH}). Following identical algebra to that following (\ref{abcorrrob}), one finds that the Bell inequality is violated.  Obviously in this scenario, the Bell inequality is violated not because there is entanglement. Here the demon has information about potentially non-local information of the measurement outcomes at $ A $ and $ B $, and the use of the formula (\ref{demonprob}) simply mimics the probabilities of the Bell measurement that would be performed in Rob's frame.  

The only difference between Fig. \ref{fig1}(b) and the demon example is that entanglement is used to produce the probabilities (\ref{robprobs}).  The role of entanglement in Rob's frame is to produce the duplicate results at $CD$ as given in Eq. (\ref{intermediatestaterob}).  The entanglement replaces the non-local powers of the demon, which bridges the distance between $ A $ and $ B$.  The question is then whether it is appropriate to interpret  Fig. \ref{fig1}(b) as verification of entanglement between $ A$ and $ B $.  In our opinion, this situation seems better described as ``entanglement-assisted fictitious violation of a Bell inequality'', rather than genuine entanglement detection.  

The puzzling aspect of this is that the only difference is a change of observer, and the experiment itself is identical. Thus if the order of the measurements were physically being reversed, one might be able to explain the discrepancy by claiming that they are fundamentally different experiments.  However, since they are different only due to a change of observer, it appears more important in this case to understand what the correct interpretation is.  

One might argue that there is in fact no entanglement in Lara's frame.  In Lara's frame, the Bell states at $ AB $ emerge completely randomly, and are statistically equivalent to a completely mixed state. In such an interpretation, the Bell violation then should occur due to a similar procedure as the demon --- the random verification measurements of the state (\ref{completelymixedstate}) are manipulated into a fictitious Bell violation. However, this would require the outcomes of the verification measurements $ l_A, l_B $ being suitably correlated to the Bell measurement outcomes $ i, j $, which cannot be produced by a completely mixed state.  It therefore suggests that the procedure as given in Sec. \ref{sec:laraframe} is necessary to produce the necessary correlations.  

From these observations it appears that the existence of the entanglement of the wavefunction is not universally agreed upon by different observers \cite{adami}.  That is, much like the notion of relativity of space and time itself, the entanglement of the wavefunction is also a relative quantity that is observer-dependent.  The fact that relativistic effects can effect entanglement has been observed in several works \cite{gingrich2002quantum,alsing2002lorentz,adami,byrnes2017lorentz}. However these cases are somewhat different in that there exists a transformation law that can convert quantities between frames. In this case, due to the quantum measurements, there is no such transformation that can convert the quantum states.  This point was also observed in other contexts \cite{aharanov1981,peres2002}, where relativistic effects were combined with quantum measurements.



\begin{figure}[t]
\includegraphics[width=\linewidth]{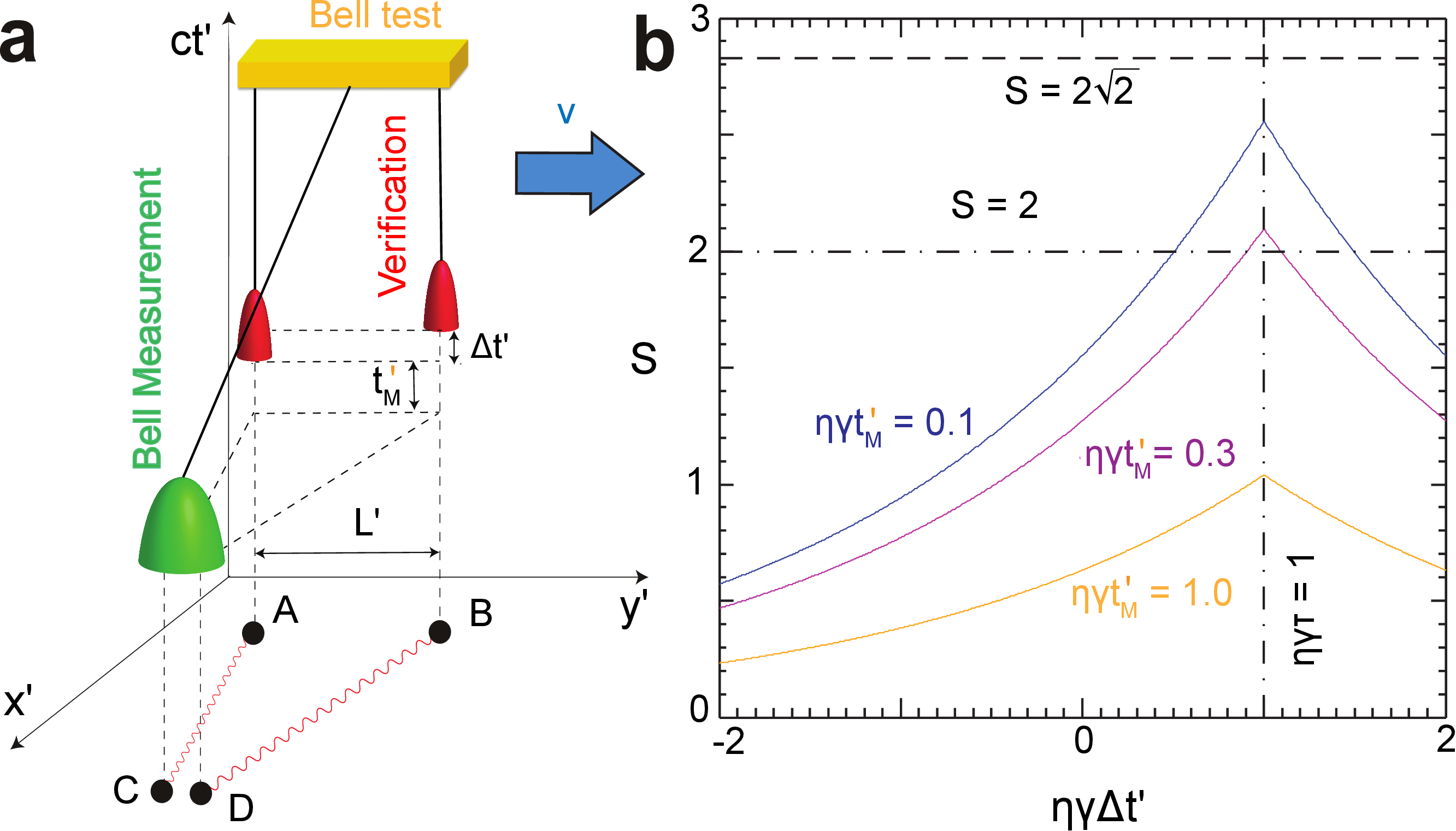}
\caption{Entanglement swapping showing entanglement in space-time (Scenario 2).  (a) The entanglement swapping protocol of Fig.~\ref{fig1}(a), according to Rob's frame who is moving in the $+y$ direction relative to Lara's frame.  The Bell measurement for Lara acts on qubits $C$ and $D$ simultaneously, which corresponds to a non-simultaneous Bell measurement with time difference $\Delta t ' $ for Rob.  The time difference between the verification measurements are adjustable and have a time difference of $\Delta t'$ in Rob's frame. (b) Bell violations for measurement times $ t'_M $ as marked in units of the depolarizing rate $ \eta $, for qubit separations such that $ \eta \gamma \tau' = 1 $.  }
\label{fig2}
\end{figure}

\section{Scenario 2: Bell correlations in space-time}

The same entanglement swapping setup can illustrate another effect, by considering boosts in other directions.  Now consider that Rob is moving in the $+y $ direction, and he is in control of the verification measurement such that 
there is a time offset $ \Delta t' $  between them.  The Bell measurement is performed in the laboratory frame (Lara) as before, and the two verification measurements occur at a time $ t_M' \pm \Delta t'/2 $, where $ t_M' $ is the midpoint between the two times. Dashed quantities refer to Rob's frame and undashed for Lara's throughout.  We also consider that some decoherence in the form of a depolarizing channel
\begin{align}
\rho_{AB} \rightarrow e^{-\eta t} \rho_{AB} + \frac{I}{4}(1- e^{-\eta t})
\end{align}
is present. This could be, for example, from storing the qubits in an imperfect quantum memory. 

Working in Lara's frame, we may calculate the outcome of the Bell violation test in the following way.  The state starts in (\ref{possible_states_a}) immediately after the Bell measurement. The depolarizing channel acts for a time $ t_M- \Delta t/2 $ until qubit $ A $ is measured.   After the measurement of qubit $ A$, the the depolarizing channel again acts on the state for a time $ \Delta t $ until  
qubit $ B $ is measured. The measurements and Bell test are performed as before, and the average CHSH inequality can be evaluated (see Appendix)  to give
\begin{equation}
\label{eq:qc14}
S = \frac{1}{4}\sum_{ij} S_{ij} = 2\sqrt{2} e^{-\eta \gamma t'_M } e^{-\eta  \gamma | \Delta t' - \tau'|/2},  
\end{equation}  
where we have written the final result in terms of Rob's variables according to the transformation
\begin{align}
t_M & = \gamma t'_M,  \nonumber \\
\Delta t & =  \gamma \left(  \Delta t' - \tau' \right),
\end{align}
where $ \gamma = (1 - \beta^2)^{-1/2}$, $\beta = v/c$, $v$ is Rob's velocity relative to Lara's frame, and $ L'$ is the distance between the qubits $A$ and $B$. Here $\tau' = \beta L'/c$ is the time offset in Rob's frame between the measurements such that in Lara's frame they are simultaneous. The calculation can be performed directly from the point of view of Rob, but due to the principle that all observers should obtain the same results, the same results are obtained in the same way as before. Interestingly, we note that Rob can actually use (\ref{eq:qc14}) in order to determine his own relative velocity, by maximizing his observed Bell violation. 

The results are illustrated in Fig. \ref{fig2}(b). If there is no decoherence ($\eta = 0 $) as can be deduced from~(\ref{eq:qc14}), one recovers the maximal Bell violations of situation in Fig.~\ref{fig1}. In the presence of decoherence, we see that in Rob's frame $ \Delta t' $ is optimal for Rob's verification measurement when $ \Delta t' =\tau'  $. This equality occurs at the particular time when the Bell verification measurement is performed simultaneously in Lara's frame. This result is natural from the point of view of Lara's frame, as any other $ \Delta t' $ would correspond to measuring the CHSH correlations at different times.  From the point of view of Rob, due to a different notion of simultaneity, he must deliberately offset his times in order to get the maximum Bell violation. The time that he must offset his time is precisely such that in Lara's frame, the measurements are carried out simultaneously.  In the limit of very strong decoherence, the Bell violations would only occur in a very narrow window of measurements centered around the time $ \Delta t' = \tau ' $.  
Rob would then conclude that due to the optimal time offset for his measurements, there are  Bell correlations both non-locally as well as non-simultaneously.  This has similarities with previous works examining bounds on entanglement in time \cite{legget1985,olson2012,budroni2013,zukowski2014}, illustrating that nonlocal quantum correlations can be engineered by measurements separated in time. As we have shown here, these correlations have influence not only in a non-local fashion, but more generally across space-time.  


\section{Summary and conclusions}

In this paper we have presented two quantum thought experiments, which highlight the peculiar nature of entanglement when relativity is involved. In the first scenario, we considered transformations such that the order of the Bell measurement and the verifications were reversed. Depending upon the observer, the interpretations of the same experiment are radically different.  In one frame, entanglement swapping is completed deterministically and entanglement between qubits $A$ and $B$ is accordingly verified. In the other frame, random outcomes of the verification measurement are manipulated into a Bell violation with classical feed-forward and no entanglement between qubits $A$ and $B$ is present at any time. The different interpretations mean that different observers disagree on whether the wavefunction is entangled or not. This example points to the possibility that quantum entanglement being a relative quantity, much like other concepts such as simultaneity in special relativity.  In the second scenario, the phenomenon of entanglement across space-time can be observed, where the maximum Bell violation occurs for measurements at different times in a suitable reference frame. In this case the effect occurs because the Bell correlations are best observed when the verification measurements are made in the same reference frame as the entangled state itself.  By preparing the state via entanglement swapping, the state is projected in the reference frame of the Bell measurement.  

It is now widely recognized that entanglement is a resource for performing useful quantum information tasks, such as quantum computing, quantum teleportation, and quantum cryptography.  But what is a physical resource? A key feature of a physical resource is
a principle of invariance such that it can be quantified in a basis-independent (i.e. observer-independent) way.  In this paper, we have shown that quantum entanglement of a two-particle state may be an observer-dependent quantity. How can something that disappears for one observer --- and then reappears for another observer --- possibly be a physical resource? Entanglement is a mathematical statement about the separability of two quantum states.  A mathematical statement cannot be a physical resource, any more than, say, the Pythagorean theorem.  This suggests that the physical resources are in fact the non-local Bell violating correlations that entanglement seems to encode in one frame, but not in another, an idea which was suggested independently in other works \cite{anders2009, raussendorf2013, abramsky2017}. It is these non-local correlations --- which are present in both frames --- that are the true physical resources.  Entanglement is simply one bookkeeping device to keep track of them.

\acknowledgements
T. B. would like to acknowledge support by the Shanghai Research Challenge Fund; New York University Global Seed Grants for Collaborative Research; National Natural Science Foundation of China (61571301); the Thousand Talents Program for Distinguished Young Scholars (D1210036A); and the NSFC Research Fund for International Young Scientists (11650110425); NYU-ECNU Institute of Physics at NYU Shanghai; the Science and Technology Commission of Shanghai Municipality (17ZR1443600). This work is dedicated to the memory of Dr. Larry Livingstone. J.P.D. would like to acknowledge support from the Air Force Office of Scientific Research, the Army Research Office, the Defense Advanced Projects Agency, the National Science Foundation, and the Northrop Grumman Corportation. P.P.R. is funded by an ARC Future Fellowship (project FT160100397). E.O.I-O. acknowledges the Talented Young Scientists Program (NGA-16-001) supported by the Ministry of Science and Technology of China.

\appendix

\section{Derivation of Eq. (\ref{CHSH})}

We derive the generalized expression for the CHSH correlations given by (\ref{CHSH}). The well-known expression for the state $ | \Psi_{00} \rangle = ( | 0 \rangle | 0 \rangle + | 1 \rangle | 1 \rangle )/\sqrt{2} $ reads
\begin{align}
S_{00} = \sum_{n=0}^1 \sum_{m=0}^1 \bra{\psi_{00}} (-1)^{(1-n)m} \hat{\mathcal{A}}_n \hat{\mathcal{B}}_{m}  \ket{\psi_{00} }.
\label{s00expression}
\end{align}
To obtain the CHSH correlations for the other Bell states, we introduce the operator
\begin{align}
U_{ij} = (\sigma^x_B)^i (\sigma^z_B)^j  
\end{align}
which satisfies
\begin{align}
U_{ij}^\dagger U_{ij} = I .
\end{align}
Introducing factors of unity before and after the operators, the corresponding correlation reads
\begin{align}
S_{ij} & = \sum_{n=0}^1 \sum_{m=0}^1 (-1)^{(1-n)m} \bra{\psi_{00}} U_{ij}^\dagger U_{ij}  \hat{\mathcal{A}}_n 
\hat{\mathcal{B}}_{m} U_{ij}^\dagger U_{ij}   \ket{\psi_{00} } \nonumber \\
& = \sum_{n=0}^1 \sum_{m=0}^1 (-1)^{(1-n)m} \bra{\psi_{ij}}  U_{ij}  \hat{\mathcal{A}}_n \hat{\mathcal{B}}_{m}  U_{ij}^\dagger  \ket{\psi_{ij} } .
\label{sijexp}
\end{align}
The transformation only acts on the $ \hat{\mathcal{B}}_{m} $ operator and can be evaluated as
\begin{align}
U_{ij} \hat{\mathcal{B}}_m U^\dagger_{ij} & = (-1)^j \left[ \frac{ \sigma_B^x +(-1)^{m+i+j} \sigma_B^z}{\sqrt{2}}  \right]   \nonumber \\
& = (-1)^j  \hat{\mathcal{B}}_{m+i+j}.
\end{align}
Substituting this into (\ref{sijexp}) we obtain
\begin{align}
S_{ij} = \sum_{n=0}^1 \sum_{m=0}^1 \bra{\psi_{ij}} (-1)^{(1-n)m+j} \hat{\mathcal{A}}_n \hat{\mathcal{B}}_{m+i+j}  \ket{\psi_{ij} },
\label{CHSHapp}
\end{align}
which is Eq. (\ref{CHSH}).

\section{Derivation of Eq. (\ref{eq:qc14})}

We consider that initially the state is prepared in the state given (1).  At $ t= 0 $, the Bell measurement in Lara's frame collapses the state to 
\begin{align}
\rho_0 = | \psi_{ij} \rangle \langle \psi_{ij} |
\end{align}
The depolarizing channel subsequently acts on this state until Rob performs the verification measurement.  The time evolution of the state before the first verification measurement is given by
\begin{equation}
\label{eq:sup02}
\rho_{AB}(t) = e^{-\eta t}\rho_0 + (1- e^{-\eta t})\frac{{I}}{4},
\end{equation}
Let us first assume that $ \Delta t > 0 $.  At the time $t_M - \Delta t/2$ in Lara's frame, Rob performs a measurement on the state $\rho_{AB}(t)$ for the qubit at $A$. 
We note that since $\hat{\mathcal{A}}_n$ and $\hat{\mathcal{B}}_m$ act on different Hilbert spaces, $\hat{\mathcal{A}}_n$ and $\hat{\mathcal{B}}_m$ commute. As such, the order of CHSH verification measurement is irrelevant. The state immediately after the first verification measurement becomes
\begin{equation}
\label{eq:sup03}
\sigma_{AB}^{(n)} (t_M  - \Delta t/2) = \sum_{l_A} \Pi_A^{(n)} (l_A)  \rho_{AB}(t_M -\Delta t/2) \Pi_A^{(n)} (l_A) ,
\end{equation} 
where 
\begin{align}
\Pi_A^{(n)} (l_A)  = \lvert  l_A \rangle^{(n)} \langle  l_A \rvert^{(n)}  \otimes {I}_B \otimes {I}_{C} \otimes {I}_{D} 
\end{align}
is the measurement operator on qubit $ A $. 
This state again evolves in time due to the depolarizing channel as
\begin{align}
 \sigma'^{(n)}_{AB} (\tilde{t}) &  = e^{-\eta \tilde{t}} \sigma_{AB}^{(n)} (t_M  - \Delta t/2) 
+ (1 - e^{-\eta \tilde{t}})\frac{{I}}{4}, \nonumber \\
& =  \sum_{l_A} \Pi_A^{(n)}(l_A)  \Big[ e^{-\eta \tilde{t}}  \rho_{AB} (t_M -\Delta t/2) \nonumber \\ 
& + (1 - e^{-\eta \tilde{t}})\frac{{I}}{4} \Big]  \Pi_A^{(n)} (l_A)
\label{eq:sup04}
\end{align}
where $\tilde{t} $ is the time after the measurement on qubit $ A $.  This then further evolves until Rob performs a measurement on qubit $B$. Similarly, at time $\tilde{t} = \Delta t$, Rob makes a measurement with operator 
\begin{align}
\Pi_B^{(m)} (l_B) = {I}_A \otimes  \lvert l_B\rangle^{(m)} \langle l_B\rvert^{(m)} \otimes {I}_{C} \otimes {I}_{D}
\end{align}
on the particles at $B$, the state $\sigma'^{(n)}_{AB}(\tilde{t}= \Delta t)$ transforms as 
\begin{align}
\label{eq:sup05}
\sigma''^{(n,m)}_{AB}(\Delta t) &  = \sum_{l_B}  \Pi_B^{(m)} (l_B) \sigma'^{(n)}_{AB}(\Delta t) \Pi_B^{(m)} (l_B) \nonumber \\
&=  \sum_{l_A l_B}  \Pi_A^{(n)} (l_A) \Pi_B^{(m)} (l_B) \rho_{AB} (t_M +\Delta t/2)  \nonumber \\
& \times \Pi_B^{(m)} (l_B) \Pi_A^{(n)} (l_A).
\end{align}
The second form of the equation shows that the state is equivalent to simply evolving the state to a time $ t_M +\Delta t/2 $ under the depolarizing channel, and then performing a measurement. If the order of the measurements were reversed (i.e. $ \Delta t< 0  $) we would follow the same procedure and find that the final state is the measurement at the later time $ t_M -\Delta t/2 $.  Hence in general the procedure is equivalent to taking the expectation value with respect to the state  $  \rho_{AB} (t_M +| \Delta t |/2) $.  

The above reasoning can be used to evaluate the expectation value as in (\ref{CHSH}). For a given $ i,j $ outcome we have 
\begin{align}
\label{eq:sup06}
S_{ij} & = (-1)^j\sum_{n,m = 0}\mathrm{Tr}\left(\hat{\mathcal{A}}_n \hat{\mathcal{B}}_{m+i+j}  \rho_{AB} (t_M + |\Delta t |/2) \right) \nonumber\\
 & = \frac{e^{-\eta(t_M - |\Delta t |/2)}e^{-\eta |\Delta t |}}{\sqrt{2}} \sum_{n,m = 0}^1 (-1)^{2[(m+ i)(1-n) + j]},\nonumber\\
& = 4 \frac{e^{-\eta(t_M - |\Delta t |/2)}e^{-\eta |\Delta t |}}{\sqrt{2}}.
\end{align}
Taking an average over all outcomes, we obtain
\begin{equation}
\label{eq:sup07}
S = \frac{1}{4} \sum_{i,j} S_{i,j} = 2\sqrt{2} {e^{-\eta(t_M +|\Delta t|/2)}}.
\end{equation}

To relate time $t$ in Lara's frame to the time $t'$ in Rob's frame, we use the standard relativistic transformation
\begin{align}
t = \gamma (t' - \beta \tfrac{x'}{c}),
\end{align}
where $\beta = \tfrac{v}{c}$, $v$ is Rob's velocity relative to Lara's frame, $c$ is the speed of light, and $\gamma = (1 - \beta^2)^{-1/2}$. The time difference $\Delta t $ between the measurement of $A$ and $B$ in Lara's frame transforms as  
\begin{align}
\Delta t = \gamma\left( \Delta t' - \tau'\right),
\label{delttrans}
\end{align}
where
\begin{align}
\tau' =  \frac{\beta L'}{c} .
\end{align}
Suppose that the average time $t_M$ of measurement in Lara's frame is $t_M = (t_A + t_B)/2$ where $t_{i}$, ($i= A,B$) is the time to make the measurement on particles at $A$ and $B$.  Then, $t_M$ would transform as
\begin{equation}
\label{eq:sup08}
t_M = \gamma\left( \frac{t'_A + t'_B}{2}  - \beta\frac{2x'_A + L'}{2c}\right).
\end{equation}
We then have 
\begin{align}
t_M' = \frac{t'_A + t'_B}{2}  - \beta \frac{2x'_A + L'}{2c},
\end{align}
so that 
\begin{align}
t_M = \gamma t'_M .
\label{tmtrans}
\end{align}
Substituting (\ref{delttrans}) and (\ref{tmtrans}) into (\ref{eq:sup07}) gives (\ref{eq:qc14}).


\end{document}